\documentclass[prl,showpacs,twocolumn,preprintnumbers,amsmath,amssymb,superscriptaddress]{revtex4}

\usepackage{graphicx}
\usepackage{dcolumn}
\usepackage{bm}

\begin{document}
\title{
Line Broadening and Decoherence of Electron Spins in Phosphorus-Doped Silicon
Due to Environmental $^{29}$Si Nuclear Spins}
\author{Eisuke Abe}
\affiliation{
Department of Applied Physics and Physico-Informatics,
Keio University, 3-14-1 Hiyoshi, Yokohama 223-8522, Japan}
\author{Akira Fujimoto}
\affiliation{
CREST, Japan Science and Technology Agency,
4-1-8 Honcho, Kawaguchi, 332-0012}
\author{Junichi Isoya}
\affiliation{
Research Center for Knowledge Communities, University of Tsukuba,
1-2 Kasuga, Tsukuba City 305-8550, Japan}
\author{Satoshi Yamasaki}
\affiliation{
Diamond Research Center,
National Institute of Advanced Industrial Science and Technology,
Tsukuba Central 2, 1-1-1 Umezono, Tsukuba City 305-8568, Japan}
\author{Kohei M. Itoh}
\email{kitoh@appi.keio.ac.jp}
\affiliation{
Department of Applied Physics and Physico-Informatics,
Keio University, 3-14-1 Hiyoshi, Yokohama 223-8522, Japan}
\affiliation{
CREST, Japan Science and Technology Agency,
4-1-8 Honcho, Kawaguchi, 332-0012}
\date{\today}
\begin{abstract}
Phosphorus-doped silicon single crystals with 0.19 \% $\leq f \leq$ 99.2 \%,
where $f$ is the concentration of $^{29}$Si isotopes,
are measured at 8 K using a pulsed electron spin resonance technique,
thereby the effect of environmental $^{29}$Si nuclear spins
on the donor electron spin is systematically studied.
The linewidth as a function of $f$ shows a good agreement with theoretical analysis.
We also report the phase memory time $T_M$ of the donor electron spin
dependent on both $f$ and the crystal axis relative to the external magnetic field.
\end{abstract}
\maketitle
Electron spin resonance (ESR) in phosphorus-doped silicon (Si:P)
at low temperatures is now a textbook example of
how an electron spin in a solid interacts with nuclei.
In this system, the donor $^{31}$P nucleus,
the electron bound to it,
and $^{29}$Si nuclei randomly occupying
$f$ = 4.7 \% of the lattice sites of natural Si,
all carry the smallest possible spin angular momentum $\hbar$/2.
The ESR spectrum exhibits a doublet separated by 4.2 mT
due to the contact hyperfine (hf) interaction with $^{31}$P.
In addition, the electron is subject to the hf field from $^{29}$Si nuclei,
since the orbital part of the electron wavefunction
spreads over thousands of lattice sites.
It was soon realized that this interaction inhomogeneously broadens the ESR lines.
A $^{29}$Si nucleus at the $l$th site shifts
the precession frequency of the electron spin
by $\pm a_l/2$, according to the direction of the nuclear spin.
$a_l$ is the isotropic hf constant at the $l$th site.
As long as the number of the electron spins participating in the ESR line is large
and the total amount of the hf field each electron spin feels is stochastic,
the resultant line shape will be approximated by the Gaussian distribution,
for which the FWHM linewidth $ \Delta B_e$ and
the root-mean-square linewidth $\Delta B_{\mathrm{rms}}$ are related by
\begin{equation}
    \Delta B_e = 2 \sqrt{2 \ln 2} \ \Delta B_{\mathrm{rms}}.
\label{fwhm-rms}
\end{equation}
$\Delta B_{\mathrm{rms}}$ is written,
neglecting the polarization of $^{29}$Si nuclear spins, as~\cite{Koh57}
\begin{equation}
    \Delta B_{\mathrm{rms}}  = \Big[ f \sum_l \big( a_l/2 \big)^2 \Big]^{1/2},
\label{rms}
\end{equation}
where $a_l$ is rescaled in the unit of tesla,
and the sum runs over all the sites but the origin.
Electron-nuclear double resonance (ENDOR)
has allowed the experimental determination of $a_l$ at a number of sites, 
and a good agreement was found between the linewidth in natural Si
and the one deduced from Eq.~(\ref{rms})~\cite{Feh59}.
However, neither is obvious, when $f$ is reduced,
if the square root dependence on $f$ still holds
nor the line remains Gaussian.
This is one of the subjects we will verify in this contribution.

The inhomogeneous line broadening is a consequence of
\textit{static} shifts of the electron spin precession frequencies.
In reality, $^{29}$Si nuclei which caused the shift are
mutually coupled through the magnetic dipolar interactions.
Hence the nuclei are also able to cause
a \textit{temporal} change of the electron spin precession frequency by flip-flopping their spins.
This leads to decoherence of the electron spins due to spectral diffusion~\cite{HH56},
which has gained revived attention in connection with spin-based quantum computing.
It is suggested that decoherence due to nuclear-induced spectral diffusion
could be an ultimate bottleneck for quantum computing schemes
based on GaAs quantum dots (QDs)~\cite{dSDS03,WdSDS05,YLS05}.
In this context, Si:P is a promising testbed for studying nuclear-induced spectral diffusion,
not only because quantum computers based on Si:P have been proposed~\cite{Kan98},
but also because the concentration of environmental nuclei is controllable,
which is impossible in III-V compounds whose available isotopes all possess non-zero nuclear spins.

In this contribution, we experimentally study
inhomogeneous line broadening and
nuclear-induced decoherence of the P donor electron spins
in isotopically controlled Si single crystals
whose $^{29}$Si concentrations $f$ range from  0.19 \% to 99.2 \%.
Isotopically purified Si is now in wide use for the study of isotope effects in Si~\cite{CT05},
and preparations of the samples with controlled $f$ (such as 50 \%)
were accomplished in the following manner:
First, two bars of isotopically purified $^{29}$Si or $^{28}$Si and natural Si were set together,
and partially melted into one by optical heating.
The mass ratio of the two bars were chosen to achieve desired $f$.
The combined bar was made single-crystal by the floating zone (Fz) method
using another bar of natural Si as a seed crystal.
All the procedures were carried out in high purity argon gas atmosphere.
Care was taken to achieve uniform distribution of $^{29}$Si atoms.
Finally, the single crystals of about 5 mm in diameter
and a few centimeters in length were grown along [001].
Information on the samples used in this work is summarized in Table~\ref{sample}.
The net donor concentrations $N_d$ were controlled to $\sim$10$^{15}$ cm$^{-3}$
to minimize unwanted dipolar interactions among the donor electron spins,
but still ensuring the measurement sensitivity.
\begin{table}
\caption{\label{sample}
Summary of isotopically controlled Si single crystals.
The $^{29}$Si concentration $f$ [\%],
crystal growth method (Czochralski or floating zone),
net donor concentration $N_d$ [10$^{15}$ cm$^{-3}$],
spin-lattice relaxation time $T_1$ [ms],
FWHM of the ESR lines $\Delta B_e$ [mT], and
$R \equiv M_4/3(M_2)^2$, obtained from
experiments ($R_{\mathrm{exp}}$) and
simulations ($R_{\mathrm{sim}}$), are listed.}
\begin{ruledtabular}
\begin{tabular}{ldccddcc}
\multicolumn{1}{c}{Sample}&
\multicolumn{1}{c}{$f$\footnotemark[1]}&
\multicolumn{1}{c}{Cz/Fz\footnotemark[2]}&
$N_d$\footnotemark[3]&
\multicolumn{1}{c}{$T_1$}&
\multicolumn{1}{c}{$\Delta B_{e}$}&
$R_{\mathrm{exp}}$&
$R_{\mathrm{sim}}$ \\
\hline
$^{29}$Si-0.2\%\footnotemark[4] & 0.19 & Cz & 0.3 & 11.0 & 0.023 & 1.9 & $-$\footnotemark[6]\\
$^{29}$Si-0.3\% & 0.31 & Fz & 0.6 & 13.1 & 0.082 & 1.7 & $-$\footnotemark[6] \\
$^{29}$Si-1\% & 1.2 & Fz & 0.7 & 10.0 & 0.078 & 1.6 & 1.65 \\
$^{29}$Si-5\%\footnotemark[5] & 4.7 & Cz & 0.8 & 15.8 & 0.26 & 1.0 & 1.15 \\
$^{29}$Si-10\% & 10.3 & Fz & 1.6 & 14.6 & 0.42  & 1.0 &1.06 \\
$^{29}$Si-50\% & 47.9 & Fz & 0.6 & 14.9 & 0.89 & 1.0 & 1.00 \\
$^{29}$Si-100\%\footnotemark[4]\footnotemark[5] & 99.2 & Cz & 0.8 & 4.4 & 1.22 & 1.1& 0.98 \\
\end{tabular}
\end{ruledtabular}
\footnotetext[1]{Determined by secondary ion mass spectroscopy (SIMS).}
\footnotetext[2]{Crystal growth by the Fz method was carried out
with the same apparatus as used in Ref.~\onlinecite{TIO+99}.}
\footnotetext[3]{Determined from the temperature dependent Hall effect
measured by the van der Pauw method (down to about 20 K).
Residual acceptor concentrations are of the order of 10$^{13}$-10$^{14}$ cm$^{-3}$.}
\footnotetext[4]{Further information is found in Ref.~\onlinecite{IKU+03}.
$^{29}$Si-0.2\% here corresponds to isotopically purified $^{30}$Si in the reference.}
\footnotetext[5]{The same samples as used in our previous work (Ref.~\onlinecite{AIIY04}).}
\footnotetext[6]{The simulations exhibit significant counts at the origin
(no nuclei are found in any of the 176 sites), and $R$ is not calculated.}
\end{table}

Pulsed ESR experiments were carried out
using a Bruker Elexsys E580 spectrometer
equipped with an Oxford ER4118CF cryostat.
We exclusively applied a Hahn echo sequence
given by $\pi$/2-$\tau$-$\pi$-$\tau$-Echo,
where $\tau$ is the interpulse delay
and the duration of the $\pi$/2 pulse was set to 16 ns.
The measurement temperature
must be chosen properly~\cite{AIIY04}:
It has to be below 10 K for the study of spectral diffusion,
otherwise the phase memory time $T_M$ is dominated by the spin-lattice relaxation time $T_1$
due to an Orbach process~\cite{YJK+03,TLAR03}.
On the other hand, at temperatures below 4 K,
extremely long $T_1$ reduces the efficiency of our pulsed experiments
in which each echo sequence needs to be repeated
at a time interval much longer than $T_1$.
We found 8 K to be optimal:
low enough for $T_M$ not to be limited by $T_1$ 
but high enough to ensure a reasonable measuring time.
$T_1$ listed in the fifth column of Table~\ref{sample},
measured by an inversion recovery method,
vary with samples, but show no marked correlations with $f$ nor $N_d$.
We presume that this difference in $T_1$ has little effect on the difference in $T_M$,
as long as $T_1$ of one sample is much longer than its $T_M$.
The repetition rate of the Hahn echo sequence was then set to 80 ms.

\begin{figure}
\includegraphics[scale = 0.42]{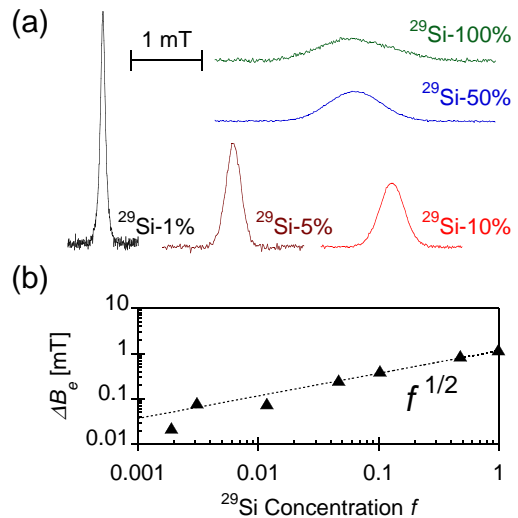}
\caption{\label{FS}(Color online)
(a) Echo-detected ESR spectra of Si:P.
Only the normalized high-field lines are shown.
The horizontal (magnetic field) and vertical (echo intensity) axes are arbitrarily shifted,
and the small background signals are subtracted for clarity.
(b) FWHM linewidth $\Delta B_e$ as a function of $f$.
The dotted line is a theoretical prediction, as explained in the text.}
\end{figure}

The echo-detected ESR spectra were obtained
by measuring the intensity of the Hahn echo
as a function of the external magnetic field $B_0$ with $\tau$ fixed.
The high-field lines of the doublets are shown in Fig.~\ref{FS}-(a).
The lines are Gaussian in $f \geq$ 4.7 \%,
while in $f \leq $ 1.2 \% the lines resembled Lorentzian (shown only for $^{29}$Si-1\%).
The same features were confirmed from the low-field lines (not shown).
Comparison between experimental linewidths and
those deduced from Eqs.~(\ref{fwhm-rms}) and (\ref{rms}) is straightforward.
Hale and Mieher have determined $a_l$ for 176 lattice sites through ENDOR~\cite{HM69},
and we will use the values reported in Table~II of Ref.~\onlinecite{HM69},
dividing them by $g_e \mu_B h$.
Figure~\ref{FS}-(b) plots both theoretical and experimental $\Delta B_e$ as a function of $f$,
which agree quite satisfactory.
$\Delta B_e$ at $f$ = 99.2 \% is calculated to be 1.15 mT,
which accounts for 94 \% of the experimental value, 1.22 mT.
The underestimation of 6 \% is acceptable, since we included only 176 sites.
The rest of thousands of sites will have very small isotropic hf constants,
and increase the linewidth further, but not largely.
This result, together with the fact that the linewidths are independent of the crystal orientations
within experimental errors, justifies the exclusion of the anisotropic term of the hf interaction.

\begin{figure}
\includegraphics[scale = 0.41]{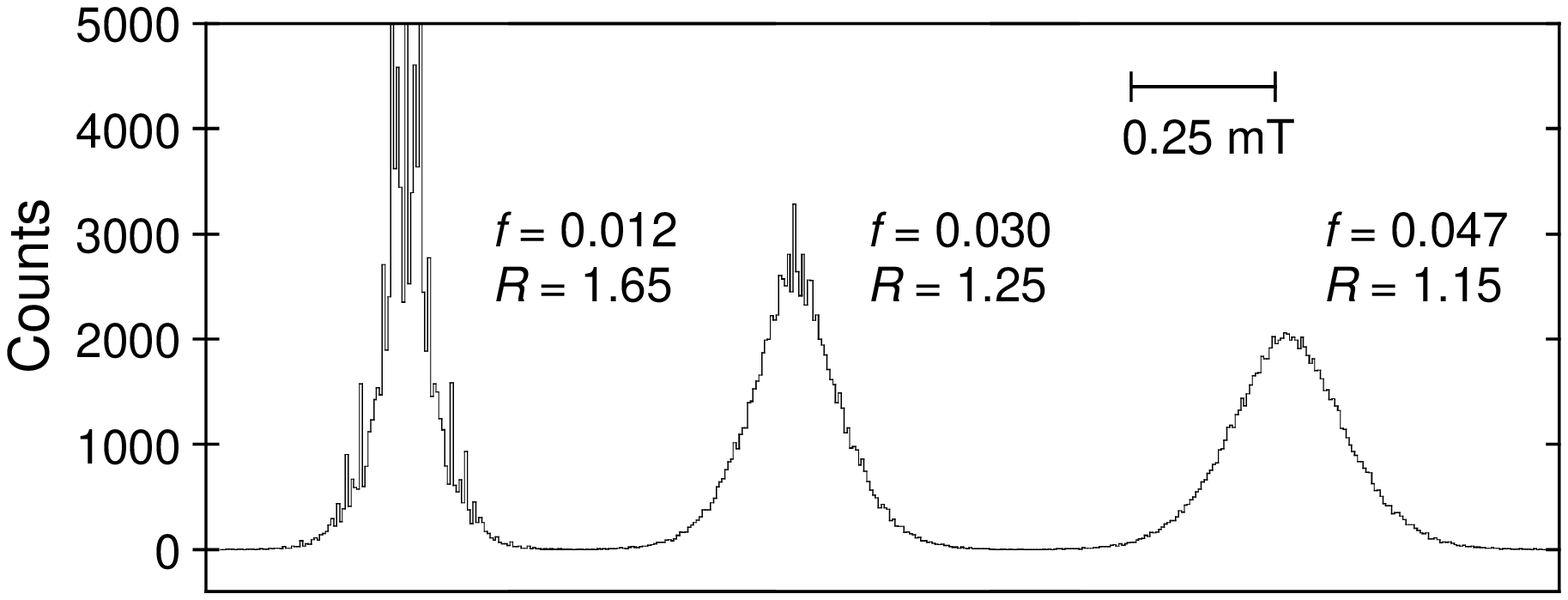}
\caption{\label{HG} Simulations of the ESR lines of Si:P
for $f$ = 1.2 \%, 3.0 \%, and 4.7 \%.
The total counts for each line is 10$^5$,
and the width of  each bar is set to 0.005 mT.
The horizontal (magnetic field) axes are arbitrarily shifted.}
\end{figure}

The line shapes in $f \leq$ 1.2 \% are no longer Gaussian,
and the use of Eq.~(\ref{fwhm-rms}) is dubious.
A general method to characterize a line shape is to
calculate the quantity $R$ defined as $M_4/3(M_2)^2$,
where $M_2$ and $M_4$ are
the second and fourth moments of the line~\cite{Abr60}.
A Gaussian curve has $R$ of unity.
On the other hand, $R$ for a pure Lorentzian curve
cannot be defined, since both $M_2$ and $M_4$ diverge.
Hence $R$ is a convenient measure to quantify the deviation from Gaussian.
In the seventh column of Table~\ref{sample},
$R$ numerically calculated from the experimental lines are given.
We set moderate cut-offs to exclude the contribution
from signals remote from the line centers~\cite{Abr60}.
$R$ deviates from unity for $f \leq$ 1.2 \%, and $R$ = 1.6 $\sim$ 1.9 is
consistent with the previous report by Feher in isotopically purified $^{28}$Si of 99.88 \% 
($R$ = 1.8 is given in Table~V of Ref.~\onlinecite{Feh59}).
The reason for this change to happen may be explained as follows:
When $f$ is reduced, it becomes rare to encounter $^{29}$Si nuclei near the donor,
where $a_l$ can be larger than the observed $\Delta B_e$~\cite{Hf}.
Once such a $^{29}$Si nucleus having a large $a_l$ is found,
the electron spin is kicked out from the center part of the line,
and there is no chance of its returning to the center,
since other $^{29}$Si nuclei are likely to have small $a_l$.
Those electron spins can only contribute to form wings of the line.
This qualitative argument can be checked by a simple simulation,
in which the known 176 sites are assumed to be occupied,
independently with probability $f$, by $^{29}$Si nuclei
which are up or down with probability one-half.
Then the hf shift of an electron spin is computed.
Repeating the procedure 10$^5$ times,
a frequency histogram should portray the observed line.
The eighth column of Table~\ref{sample} lists
$R$ calculated from each simulated line,
and Fig.~\ref{HG} shows simulated lines for $f$ = 1.2 \%, 3.0 \%, and 4.7 \%.
It is seen that $R$ gradually increases as reducing $f$,
and deviates from unity.
The agreement between $R_{\mathrm{exp}}$ and
$R_{\mathrm{sim}}$ for $f$ = 1.2 \% is good,
even though the histogram is strongly peaked at the origin
due to the limited number of the sites considered.
It is interesting to continue this work to lower-$f$ samples
in order to seek for the intrinsic linewidth,
or the boundary at which other broadening mechanisms prevail~\cite{Lw}.

\begin{figure}
\includegraphics[scale = 0.4]{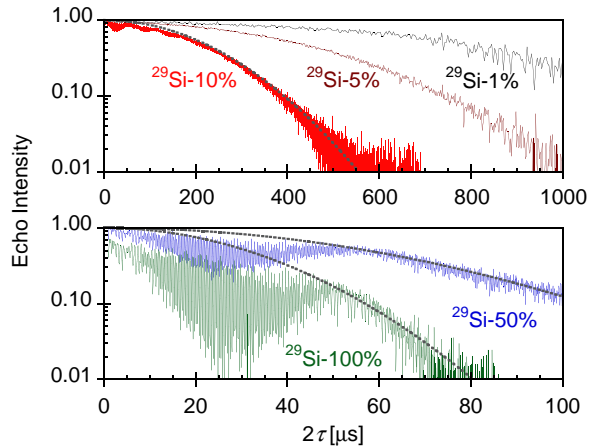}
\caption{\label{DC}(Color)
Hahn echo decay curves as a function of 2$\tau$
when $B_0$ is along the [001] crystal axis.
Note that the scales of horizontal axes in upper and lower panels
differ by an order of magnitude.
Gaussian fits to the echo decay envelopes
are superposed on the data of $^{29}$Si-10\%, 50\%, and 100\%.
The oscillations due to ESEEM are also visible.}
\end{figure}

Now we examine the echo decay behavior by changing $\tau$.
$B_0$ was fixed at the centers of the high-field lines.
We define $T_M$ as the time at which
the envelope of an echo decay curve damps to 1/$e$ of its initial echo intensity.
Figure~\ref{DC} shows echo decay curves as a function of  2$\tau$,
when $B_0$ is along the [001] crystal axis.
The strong oscillation observed in the high-$f$ samples
is electron spin echo envelope modulation (ESEEM).
Since this effect has already been analyzed by us~\cite{AIIY04}
and by Ferretti \textit{et al.}~\cite{FFPS05},
it will not be dealt in the present contribution.

\begin{figure}
\includegraphics[scale = 0.42]{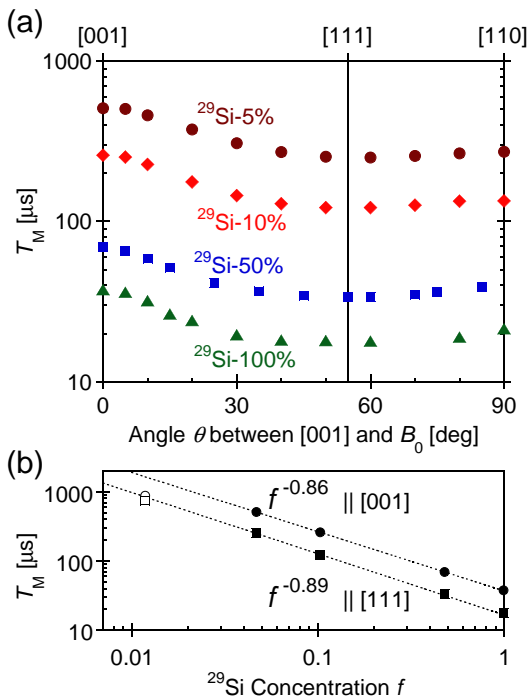}
\caption{\label{OD}(Color online)
(a) Angular and (b) $^{29}$Si concentration dependence of $T_M$.
The circles (squares) in (b) represent $T_M$ obtained when $B_0$
is along [001] (approximately [111]).
Only the data points in the filled symbols are used for the fittings.
}
\end{figure}

For samples with $f \geq$ 4.7 \%,
the echo decay envelopes are obscured by ESEEM.
To deduce $T_M$, we approximate them as Gaussian of the form $\exp[-(2\tau/T_M)^2]$.
To determine a more accurate form of the decay envelopes,
which could contain several time constants of different origins,
requires further research.
We repeat the experiments as rotating the samples about the [1$\bar{1}$0] axis
to characterize the angular dependence of $T_M$.
It takes maximum when $B_0 \parallel$ [001] and
minimum when $B_0 \parallel$ [111] as shown in Fig.~\ref{OD}-(a).
This result bridges a large gap of $f$ in our previous report of $T_M$
carried out when only $^{29}$Si-100\% and 5\% were available~\cite{AIIY04}.
$T_M$ also exhibits a power-law-like dependence on $f$ as shown in Fig.~\ref{OD}-(b). 
Empirical fittings suggest $T_M \propto f^{\gamma}$
with $\gamma$ = $-$0.86 for $B_0 \parallel$ [001] and $-$0.89 for $B_0 \parallel$ [111].
The $^{29}$Si concentration dependence confirms that $T_M$
is affected by the presence of $^{29}$Si nuclei,
and the angular dependence demonstrates that spectral diffusion plays a major role,
since the tendency reflects the strength of the nuclear dipolar couplings.
When $B_0 \parallel$ [111], one of the four tetrahedral bondings is parallel to $B_0$,
and this pair of nuclei gives rise to the strong dipolar coupling,
which makes $T_M$ shortest.
When $B_0 \parallel$ [001], all the dipolar couplings between nearest neighbors vanish,
so that $T_M$ is longest.

The angular dependence of $T_M$ in $^{29}$Si-100\% and 5\%
has been explicitly compared  with theory developed by de Sousa and Das Sarma~\cite{dSDS03}
in Fig.~1 of Ref.~\onlinecite{DSdSHK05}, showing reasonable agreement.
There are other theories on nuclear-induced spectral diffusion,
which are developed recently and applicable to Si:P as well as GaAs QDs~\cite{WdSDS05, YLS05}.
We believe our experimental results serve as a testbed for the validations of such theories.

The situation changes in $f \leq$ 1.2 \%.
The forms of the echo decay curves are quite dissimilar to Gaussian.
The fittings by a single time constant are difficult,
and $T_M$ as the $1/e$ decay points are plotted in Fig.~\ref{OD}-(b).
$T_M$ of $^{29}$Si-1\% when $B_0 \parallel$ [111]
roughly coincides with the value extrapolated
from the empirical power law.
It must be noted that we encountered
phase fluctuations of the quadrature-detected echo signals,
occurring at 2$\tau$ longer than about 500 $\mu$s,
which significantly distorted the echo decay curves.
This means, in effect,
all the echoes are generated along one predetermined direction of the rotating frame,
but each echo is detected along a random direction.
A very similar instrumental problem has also been reported by Tyryshkin \textit{et al.}
when they measured purified $^{28}$Si
($f \sim$ 0.08 \%, $N_d$ = 10$^{15}$ cm$^{-3}$)~\cite{TLAR03}.
Since our apparatus is of the same design as theirs,
we suspect the two problems arise from the same origin~\cite{Id}.
Apart from this problem, it is not surprising that decoherence mechanisms
other than spectral diffusion, such as the dipolar interactions among the electron spins
and the $T_1$ effect, both of which we have assumed to minimize,
come into play when $f$ is reduced~\cite{TLAR03,FFPS05}.
Indeed, if we extrapolate the power law to $f <$ 1 \%, $T_M$ will soon be comparable
with $T_1$ at this temperature.
Ultimately, the temperature and donor concentration dependence of $T_M$
will have to be examined in detail to fully characterize the decoherence in this realm of $f$,
which is beyond the scope of the present contribution.
Our aim here is to investigate the decoherence dominated by
nuclear-induced spectral diffusion at fixed temperature and donor concentration.

In conclusion, we prepared isotopically controlled Si single crystals,
and investigated the widths of the ESR lines and $T_M$ of the donor electron spin
as a function of the $^{29}$Si concentration $f$ (0.19 \% $\leq f \leq$  99.2 \%) at 8 K.
We confirmed the square root dependence of the linewidth in $f  \geq$ 4.7 \%,
and also analyzed the deviation seen in $f \leq$ 1.2 \%.
It is found that $T_M$ is proportional to the power of $f$ in $f \geq$ 4.7 \%,
while further research is necessary in $f \leq$ 1.2 \%.
We believe our results on $T_M$ serve as
a testbed for the validations of theories dealing with nuclear-induced spectral diffusion,
and will provide insight into other systems such as electron spins confined in QD nanostructures.

We thank H.-J. Pohl for the purified $^{29}$Si source,
and A. Takano of NTT-AT for SIMS measurements of several samples.
We also thank T. D. Ladd, S. A. Lyon, and A. M. Tyryshkin for fruitful discussions.
E. A. was supported by Japan Society for the Promotion of Science.
This work was partly supported by the Grant-in-Aid for Scientific Research
in a Priority Area ``Semiconductor Nanospintronics'' (No.14076215).

\begin{thebibliography}{20}
\bibitem[{Hf()}]{Hf}
\bibinfo{note}{According to Ref.~\onlinecite{HM69}, $a_{l}$ is 0.21 mT at the
  (004) site and other 5 equivalent sites, and 0.16 mT at the (440) site and
  other 11 equivalent sites, and so forth.}
\bibitem[{Lw()}]{Lw}
\bibinfo{note}{The reason that the linewidth of $^{29}$Si-0.3\% is broader than
  that of $^{29}$Si-1\% is not clear. We have not confirmed the presence of
  nuclear species whose concentration is comparable with that of $^{29}$Si
  ($\sim$ 10$^{20}$ cm$^{-3}$).}
\end{thebibliography}

\end{document}